\documentclass[prl,aps,twocolumn,superscriptaddress,showpacs]{revtex4-1}
\usepackage{graphicx}
\usepackage{amsmath}
\usepackage{epsfig}
\usepackage{multirow}
\usepackage{array}
\usepackage{inputenc}
\usepackage{color}
\usepackage{tabularx}
\usepackage{textcomp}
\usepackage{ulem}
\usepackage[colorlinks=true,linkcolor=black, citecolor=blue, urlcolor=blue, 
    unicode=true]{hyperref}

\begin{document}

\title{Anharmonic magnon excitations in noncollinear and charge-ordered RbFe$^{2+}$Fe$^{3+}$F$_6$}

\author{M. Songvilay}
\affiliation{School of Physics and Astronomy and Centre for Science at Extreme Conditions, University of Edinburgh, Edinburgh EH9 3FD, UK}
\author{E. E. Rodriguez}
\affiliation{Department of Chemistry and Biochemistry, University of Maryland, College Park, Maryland 20742, USA}
\author{R. Lindsay}
\affiliation{School of Physical Sciences, University of Kent, Canterbury, CT2 7NH, UK}
\author{M. A. Green}
\affiliation{School of Physical Sciences, University of Kent, Canterbury, CT2 7NH, UK}
\author{H. C. Walker}
\affiliation{ISIS Neutron and Muon Source, Rutherford Appleton Laboratory, Chilton, Didcot OX11 0QX, United Kingdom}
\author{J. A. Rodriguez-Rivera}
\affiliation{NIST Center for Neutron Research, National Institute of Standards and Technology, 100 Bureau Drive, Gaithersburg, Maryland, 20899, USA}
\affiliation{Department of Materials Science, University of Maryland, College Park, Maryland 20742, USA}
\author{C. Stock}
\affiliation{School of Physics and Astronomy and Centre for Science at Extreme Conditions, University of Edinburgh, Edinburgh EH9 3FD, UK}
\date{\today}

\begin{abstract}

RbFe$^{2+}$Fe$^{3+}$F$_6$ is an example of a charge ordered antiferromagnet where iron sites, with differing valences, are structurally separated into two interpenetrating sublattices. The low temperature magnetically ordered Fe$^{2+}$ ($S$=2) and Fe$^{3+}$ ($S$=5/2) moments form a noncollinear  orthogonal structure with the Fe$^{3+}$ site displaying a reduced static ordered moment.  Neutron spectroscopy on single crystals finds two distinct spin wave branches with a dominant coupling along the Fe$^{3+}$ chain axis ($b$-axis).  High resolution spectroscopic measurements find an intense energy and momentum broadened magnetic band of scattering bracketing a momentum-energy region where two magnon processes are kinematically allowed.   These anharmonic excitations are enhanced in this non collinear magnet owing to the orthogonal spin arrangement.

\end{abstract}

\pacs{}

\maketitle


Quasiparticles in condensed matter are generally long-lived and non-interacting with a prototypical example being magnon excitations in ordered magnetic lattices.  Classically, in the high spin limit, these excitations correspond to transverse small-angle deviations of a spin vector away from the equilibrium direction with the length of the vector remaining fixed.  This distortion of the underlying magnetic lattice is harmonic and results in underdamped spin waves.   However, the conditions under which these excitations breakdown has become important to understanding low energy properties in a variety of systems including superconductivity~\cite{Zhang1988,Anderson1987}, frustrated magnets~\cite{Paddison2016,Han2012,Vries2009}, and also quantum liquids~\cite{Woods1973,Smith1977,Fak1998,Pitaecskii1959}.   We demonstrate the breakdown of this quasiparticle notion in a classical magnet with non collinear magnetic order where spin geometry is a key ingredient establishing quasiparticle stability.

Due to enhanced phase space and also large quantum fluctuations, one-dimensional and low-spin magnets have been at the center of the search for the breakdown of conventional spin-waves into multiparticle states~\cite{Nagler1991,Tennant1993,Lake2005,Enderle2010}.  Such composite particles can be viewed as underlying bound states with fractional quantum numbers and can only be observed through decay products in scattering experiments~\cite{Wilczek1982,Zhit2013,Chernyshev2006,Chernyshev2009} due to selection rules, resulting in a momentum and energy broadened continuum cross section and renormalization~\cite{Igarashi1992,Igarashi2005,Cabrera2014} of the single-magnon dispersion and intensity.  In collinear square lattice antiferromagnets, spin wave theory predicts two-magnon processes, which are longitudinally polarized, and correspond to the simultaneous creation of two magnons of opposite signs, reducing the value of the ordered spin moment compared to the full value $S$ \cite{Christensen2007}.  The cross section scales as $1/S$~\cite{Igarashi1992} and is inherently weak in classical high-spin magnets~\cite{Huberman2005} and such processes have been generally investigated in $S=1/2$ magnets where quantum fluctuations are large.  Another means of enhancing this cross section is through a non collinear magnetic structure where longitudinal and transverse excitations are intertwined through geometry of the magnetic lattice \cite{Oh2013,Chernyshev2006,Chernyshev2009}.   In this work, we investigate such anomalous spin fluctuations in the charge ordered RbFe$^{2+}$Fe$^{3+}$F$_6$ based on an orthogonal spin geometry.  

\begin{figure}
 \includegraphics[scale=0.35]{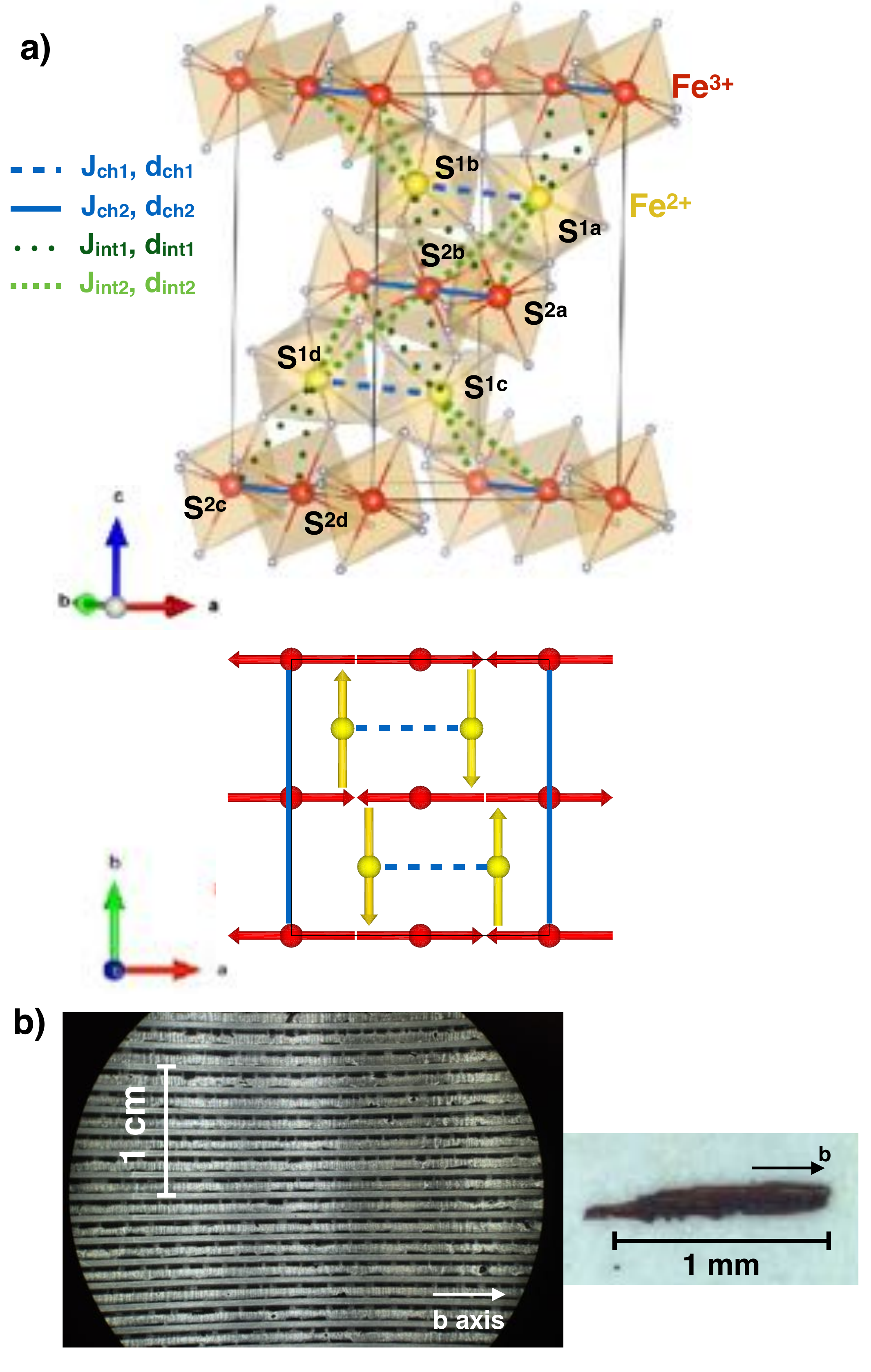}
 \caption{ \label{fig:structure}  $(a)$ Crystallographic structure of RbFe$^{2+}$Fe$^{3+}$F$_6$ with the Fe$^{3+}$F$_6$ (red) and Fe$^{2+}$F$_6$ (yellow) network and view of the perpendicular arrangement of the spins in the $(a,b)$ plane. The colored lines indicate the inequivalent nearest-neighbor exchange couplings noted J$_{ch1}$, J$_{ch2}$, J$_{int1}$ and J$_{int2}$ associated with the nearest-neighbor Fe-Fe distances $d$. \textbf{S$^{1a,b,c,d}$} and \textbf{S$^{2a,b,c,d}$} refer to each spin sites for the Fe$^{2+}$ and Fe$^{3+}$ atoms respectively. $(b)$ Photo of the sample mounting and a single crystal of RbFe$^{2+}$Fe$^{3+}$F$_6$; the small single crystals were coaligned along the $b$ axis.}
\end{figure}

RbFe$^{2+}$Fe$^{3+}$F$_6$ crystallizes in the \textit{Pnma} space group (Fig. \ref{fig:structure} $(a)$) with the lattice parameters $a=$6.9663(4), $b=$7.4390(5) and $c=$10.1216(6) \AA\ at T~=~4~K. As mentioned in \cite{Kim2012}, RbFe$^{2+}$Fe$^{3+}$F$_6$ has a structure related to the $\alpha$-pyrochlores A$_2$B$_2$X$_6$X' but with a vacancy on one of the two $A$ cations and another on the $X'$ anion site that does not contribute to the $BX_6$ octahedra.   Charge order in this compound originates from two different iron sites which have differing valences.  The RbFe$^{2+}$Fe$^{3+}$F$_6$ structure can be described as a chain of corner-shared Fe$^{3+}$F$_6$ octahedra running along $b$ and a chain of corner-shared Fe$^{2+}$F$_6$ octahedra running along the $a$-axis. The two chains are connected along the $c$ axis to form a three dimensional network. While the Fe$^{3+}$F$_6$ octahedra are only slightly distorted, a substantial distortion exists on the Fe$^{2+}$F$_6$ octahedra likely due to the Jahn-Teller effect given the underlying orbital degeneracy for octahedrally coordinated Fe$^{2+}$.~\cite{Abragram2012,Gorev2016,Molokeev2013}. 

Both magnetic iron sites order antiferromagnetically below T$_{N}$=16 K with the Fe$^{2+}$ and Fe$^{3+}$ magnetic moments oriented 90$^{\circ}$ with respect to each other forming a noncollinear structure.  As illustrated in Fig. \ref{fig:structure}  $(a)$, the Fe$^{3+}$ moments point along the $a$-axis and are coupled antiferromagnetically through nearest-neighbor interaction along the $b$-axis. The Fe$^{2+}$ moments point in the orthogonal direction ($b$-axis) and are coupled antiferromagnetically through nearest-neighbor interaction along $a$. In the low temperature ordered state, the saturated magnetic moments measured via neutron diffraction are m(Fe$^{3+}$)~=~4.29(5)$\mu_B$ (S = 5/2) and m(Fe$^{2+}$)~=3.99(5)$\mu_B$ (S = 2).  Given that the expected magnetic moment is equal to $gS$, with $g=2$ the Lande factor, while the Fe$^{2+}$ displays the full ordered magnetic moment the ordered magnetic moment on the Fe$^{3+}$ is strongly reduced. 

We apply neutron spectroscopic measurements to investigate the magnetic dynamics in RbFe$^{2+}$Fe$^{3+}$F$_6$.  Single crystals of RbFe$^{2+}$Fe$^{3+}$F$_6$ were made using hydrothermal techniques discussed in the supplementary material.  With each crystal weighing less than 1 mg, between three and five thousand were coaligned using hydrogen free Fomblin grease on a series of aluminum plates, using the long chain $b$-axis as a guide. Further measurements in the direction perpendicular to the plates indicate that this direction corresponds to the $c$-axis (see supplementary material).  The estimated total mass was between 0.3 - 0.5 g (Fig. \ref{fig:structure} $(b)$).  Neutron spectroscopy was performed using the MERLIN chopper spectrometer (ISIS, UK) and the MACS cold triple-axis (NIST, Gaithersburg) and measurements were carried out in the $(0~k~l)$ scattering plane.  Further details are supplied in the supplementary information.  The dynamical structure factor associated with the single magnon excitations was calculated using \textsc{Spinwave} \cite{Petit-SW}.

\begin{figure}
 \includegraphics[scale=0.42]{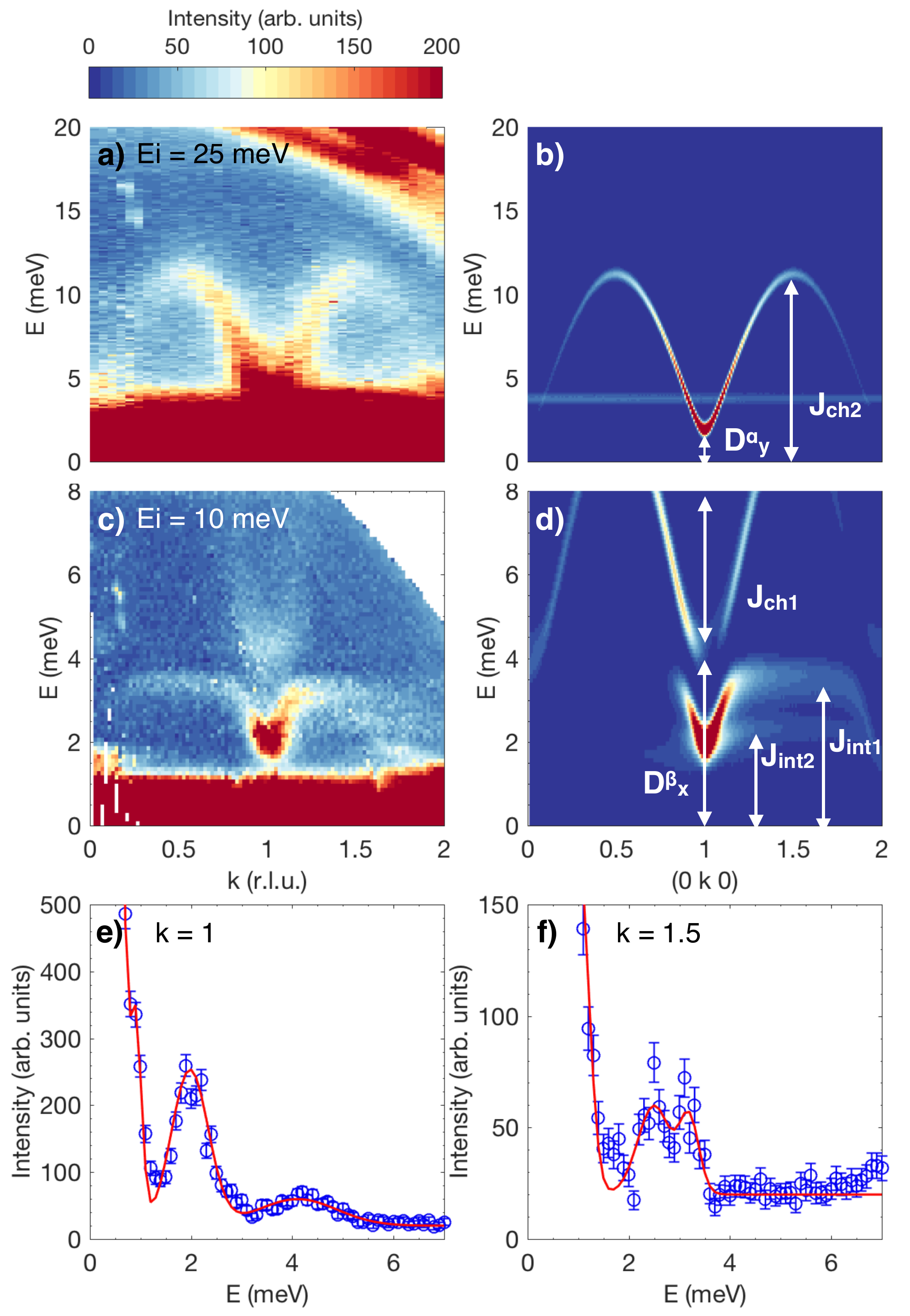}
 \caption{ \label{fig:merlin}$(a)$, $(c)$ Inelastic neutron scattering data of RbFe$_2$F$_6$ measured on MERLIN with incident energies of E$_i$ = 25 meV and E$_i$ = 10 meV at T = 5 K, along $k$. $(b)$, $(d)$ S($\mathbf{Q}$, $\omega$) simulated with spinwave calculations along (0~k~0). The white arrows indicate which band is affected by the fitted parameters. $(e)$, $(f)$ Constant-Q cuts along the k = 1 and k = 1.5 positions respectively. The red lines represent the fit of the energy position for the spinwave modes.}
\end{figure}

We first discuss the dynamics in the low temperature magnetically ordered phase measured with MERLIN.  The inelastic spectra in the $b^{*}$ direction at T~=~5~K with two incident energies of E$_i$ = 25 meV and E$_i$ = 10 meV are shown in Fig. \ref{fig:merlin} $(a)$ and \ref{fig:merlin} $(c)$ respectively. Figure \ref{fig:merlin} $(a)$ displays a spin wave dispersion stemming from $k$=1, reaching a maximum top of the band at $\sim$ 12 meV. Higher resolution data taken with E$_i$ = 10 meV displays an energy gap at the zone center of $\sim$ 2 meV, indicating an easy-axis anisotropy. Moreover, at $k$=1, the spin wave dispersion separates into two branches with a mode at 2~meV and a second one around 4 meV. The lower mode reaches a flat maximum at $\sim$ 3.5 meV while the higher mode disperses up to the 12 meV maximum observed on the E$_i$ = 25 meV data. This separation is illustrated in the constant-Q cut $k$=1 displayed in Fig. \ref{fig:merlin} $(e)$ (blue circles). In Fig. \ref{fig:merlin} $(f)$,  a constant-Q cut at $k$=1.5 shows the additional flat modes located near the magnetic zone boundary.

To understand the microscopic origin of these two modes, we have performed linear spin wave calculations based upon Heisenberg exchange and easy-axis anisotropy (described in the supplementary information). As reported in the Cs counterpart \cite{Gorev2016}, the nearest-neighbour interactions are dominated by super-exchange interactions, mediated by the F ions, requiring four different magnetic couplings.  The four couplings constants are shown in Fig. \ref{fig:structure} $(a)$ and correspond to two intrachain couplings J$_{ch1}$ and J$_{ch2}$, associated to Fe$^{3+}$-Fe$^{3+}$ and Fe$^{2+}$-Fe$^{2+}$ interactions, respectively, and two interchain couplings J$_{int1}$ and J$_{int2}$ associated to  Fe$^{3+}$-Fe$^{2+}$ interactions. 

The calculations are illustrated in Figs. \ref{fig:merlin} $(b)$ and \ref{fig:merlin} $(d)$. Panel $(b)$ illustrates a spin wave calculation where interchain interactions J$_{int1}$= J$_{int2}$=0 and panel $(d)$ shows a calculation with both inter and intra chain interactions non zero.  The band observed in the data for E$_i$ = 25 meV (Fig. \ref{fig:merlin} $(a)$ indicated by the white arrows) corresponds to the dispersion along the Fe$^{3+}$ $b$-axis chain as illustrated in panel $(b)$ which was used to adjust the J$_{ch2}$ intrachain coupling and the easy-axis anisotropy associated with this site.  The other terms of the spin Hamiltonian could be refined from the E$_{i}$=10 meV (Fig. \ref{fig:merlin} $(c)$) data. As displayed in Fig. \ref{fig:merlin} $(d)$, the white arrows indicate how the energy position of the different modes allowed the refinement of the anisotropy term for Fe$^{2+}$ and the three remaining exchange parameters.  Notably, the anisotropy term affects the position of the higher mode and the slope of this branch is also affected by J$_{ch1}$.  As for the flat modes near the zone boundary, their energy position is controlled by the interchain interactions J$_{int1}$ and J$_{int2}$.   Figures \ref{fig:merlin} $(e)$ and $(f)$ show a fit of the spin wave modes with chosen constant Q cuts at the zone center and near the zone boundary, respectively.  The best solution found for the anisotropy terms is: $D_y^{\alpha}$ = 0.075 meV (for the Fe$^3{+}$ site) and $D_x^{\beta}$ = 0.6 meV (for the Fe$^{2+}$ site) in agreement with the difference in distortion between the Fe$^{3+}$F$_6$  and Fe$^{2+}$F$_6$ octahedra. The exchange couplings extracted from the fit are J$_{ch1}$~=~1.40(5) meV (Fe$^{2+}$ chain), J$_{ch2}$~=~1.90(2) meV (Fe$^{3+}$ chain), J$_{int1}$~=~1.40(5) meV and J$_{int2}$~=~0.75(10) meV.

\begin{figure}
\includegraphics[scale=0.40]{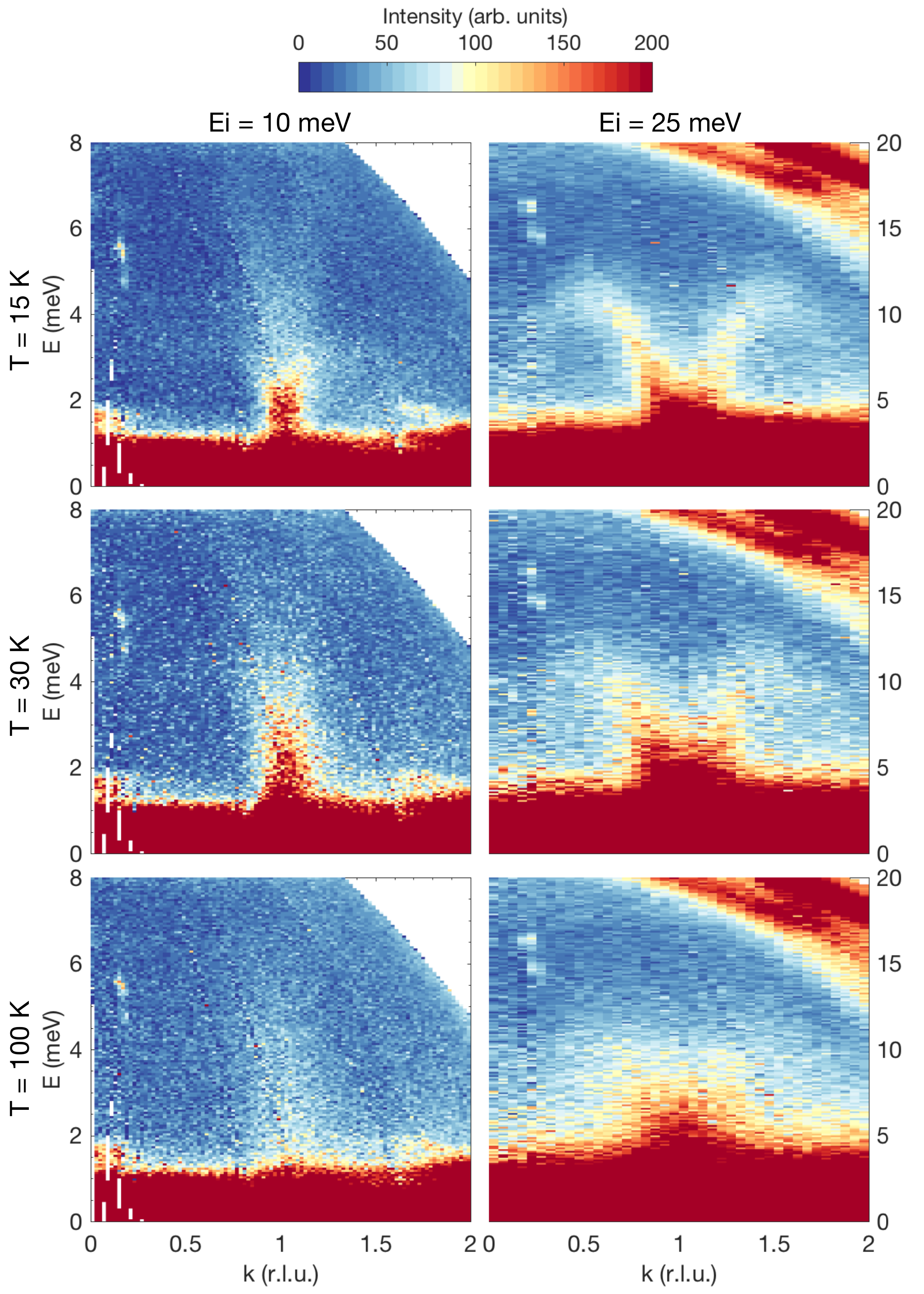}
 \caption{ \label{fig:evol_temp} Inelastic neutron scattering data of RbFe$_2$F$_6$ measured on MERLIN with incident energies of E$_i$ = 10 meV (left) and E$_i$ = 25 meV (right) along $k$ at 15 K (top), 30 K (middle) and 100 K (bottom).}
\end{figure}

The strongest coupling J$_{ch2}$  is hence found along the Fe$^{3+}$-Fe$^{3+}$ chain where the Fe-F-Fe angle is the closest to 180$^{\circ}$. Describing the system as two interacting spin chains allows an understanding of the low-energy data: without the interchain interactions, the data in the $k$ direction would only appear as a single mode stemming from $\vec{Q}$=(0 1 0) accounting for the dispersion of the Fe$^{3+}$ chain (Fig. \ref{fig:merlin} $(b)$). Because of the interaction with the Fe$^{2+}$ chain, the coupling between the two chains leads to the separation of the low-energy dispersion into two modes (Fig. \ref{fig:merlin} $(d)$).  As shown in the supplementary information, this lower energy mode is weakly dispersive along the $c$-axis while both modes show a clear dispersion along the $a$-axis, according to calculations. This indicates that the two chains seem weakly coupled despite the interchain couplings being less than an order of magnitude smaller than the intrachain couplings.  This is further confirmed by the temperature evolution of the spectra shown in Fig. \ref{fig:evol_temp}.  The interchain interaction was found to be of the order of 0.75 to 1.4 meV, which corresponds roughly to 8 to 15 K. Interestingly, near T$_N$ at T = 15 K, the low energy data shows a collapse of the two modes, giving a single branch. Hence the correlations between the two iron chains are phased out by thermal fluctuations, and the only dominant energy scale remaining is the intrachain coupling between the Fe$^{3+}$ ions. The inelastic signal also still shows a damped ``dispersion" up to ~2T$_N$$\sim$ 30~K, indicating the persistence of short range spin correlations, a behavior characteristic of low dimensional systems \cite{Huberman2008,Demmel2007}.  The persistence of short range correlations is consistent with the derived J$_{ch2}$=1.9 meV$\sim$ 22~K coupling between Fe$^{3+}$ spins.  The changes in the spin wave dispersion with temperature supports the energy scales derived from the low-temperature spin wave analysis.  

\begin{figure}
 \includegraphics[scale=0.38]{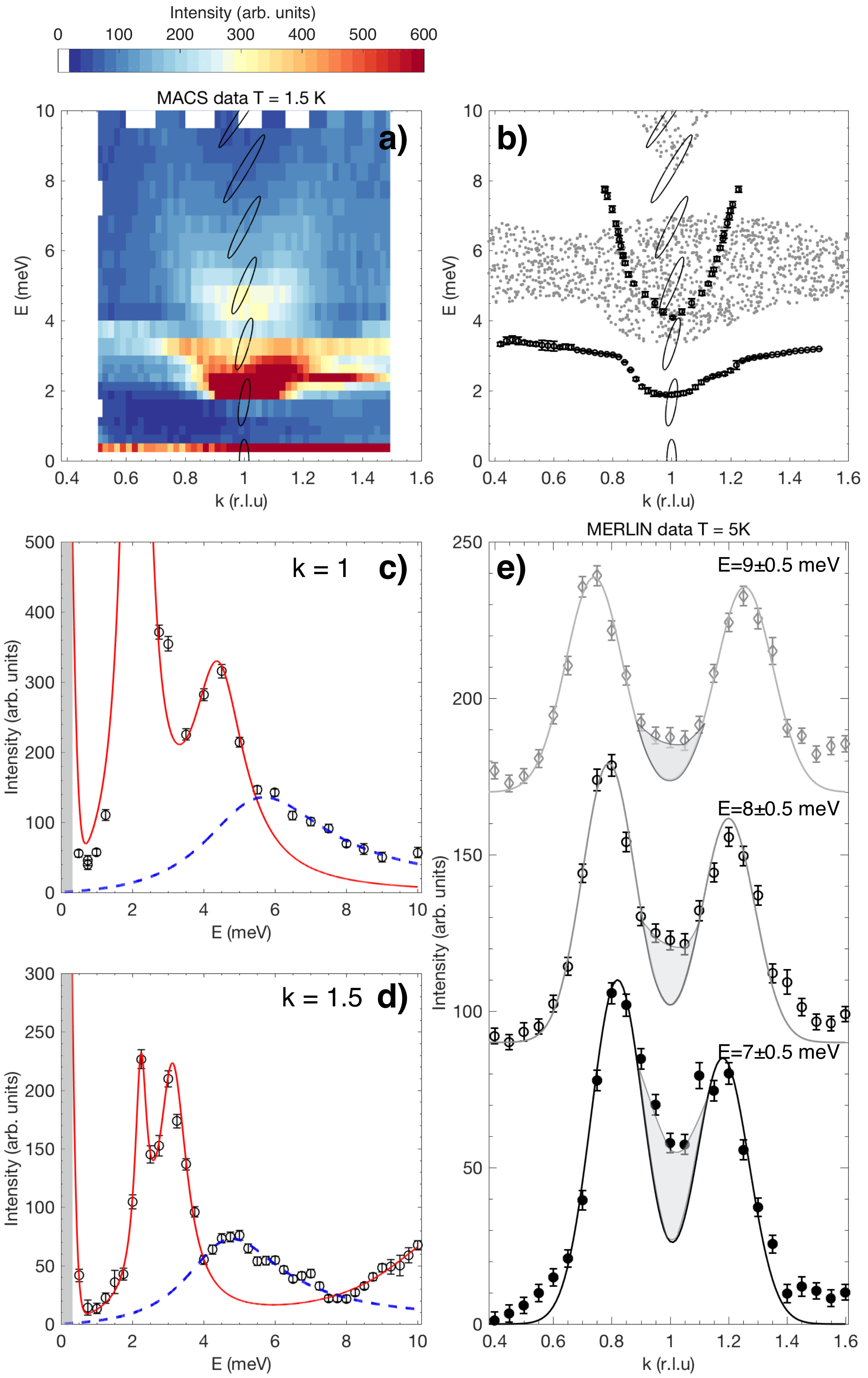}
 \caption{ \label{fig:two_magnon} $(a)$ Inelastic neutron scattering data of RbFe$_2$F$_6$ measured on MACS with final energy of E$_f$ = 3.7 meV along $k$ at 1.5 K. The grey ellipses show the evolution of the resolution ellipsoid as a function of energy along the k = 1 cut, calculated with the \textsc{Reslib} library \cite{Zheludev2007}. $(b)$ Two-magnon kinematic conditions calculated in the (E,Q) space (grey dots). The black circles are fits to the experimental data. $(c-d)$ Constant k~=1 and k = 1.5 cuts; the solid red line and blue dashed line show a fit to the data as described in the text. The grey area shows the energy resolution. $(e)$ Constant E cuts recorded on MERLIN with E$_i$ = 25 meV at 5~K.}
\end{figure} 

The noncollinear magnetic structure (illustrated in Fig. \ref{fig:structure}) brings the possibility for multi magnon states to be observable. In magnets with noncollinear spin structure, cubic anharmonic terms arise in the spin wave Hamiltonian due to the coupling of the transverse and the longitudinal fluctuations associated with deviations of the spin direction perpendicular and parallel to the ordered moment direction. These cubic terms have no analog in collinear magnets and describe the possibility of either an interaction between single and two magnons or the decay of single-magnons into pairs of other magnons \cite{Chernyshev2009, Mourigal2013}, giving rise to a continuum in the excitation spectrum. The continuum boundaries in energy and momentum are therefore determined by the single-magnon dispersion mediating such decay.  We investigate the possibility of anharmonic or multiparticle excitations in RbFe$^{2+}$Fe$^{3+}$F$_6$ in Fig. \ref{fig:two_magnon} using spectroscopy data from MACS.

Fig. \ref{fig:two_magnon} $(a)$ illustrates a color map of the excitations measured on MACS with the peak of the dispersion represented by the solid points in Fig. \ref{fig:two_magnon} $(b)$.  The grey regions in Fig. \ref{fig:two_magnon} $(b)$ (between 4 and 7 meV and between 8 and 10 meV respectively) are the regions where the Fe spins are kinematically allowed to decay, conserving both momentum and energy, given the constraints of both the low and high energy branches. In particular, the lower limit of each region is at an energy position of twice the energy position of each mode.   Fig. \ref{fig:two_magnon} $(c-d)$ show constant -$Q$ scans through the MACS data at the magnetic zone center ($k$=1) and the zone boundary ($k$=1.5).  The sharp and intense single magnon excitations (highlighted in red) are seen at low energies but also an energy broadened component with comparable integrated spectral weight is observed up to high energies of $\sim$ 10 meV (dashed blue line). This component is also extended in momentum as illustrated in panel $(e)$ and shown in panels $(c-d)$ and clearly contrasts with the sharp single magnon excitations which are resolution limited in energy and momentum.  The energy and momentum broadened cross section is not expected based on our single magnon analysis discussed above and the energy and momentum broadened nature indicates a shortened lifetime.  The region in momentum and energy where this second component of scattering is observed does coincide with the expected region based on two magnon excitations and the lower and upper branches.  Based on the broadened cross section and the comparison with calculations discussed above, we therefore conclude that this additional momentum and energy broadened component corresponds to the decay of Fe$^{3+}$ excitations into multiparticle states.

This interpretation of a decay or leakage of Fe$^{3+}$ excitations into a multiparticle continuum is also supported by magnetic diffraction data probing the magnetic structure.  Given constraints of the total moment sum rule~\cite{Hohenberg1974} of neutron scattering, the additional spectral weight appearing in the multiparticle continuum must draw from somewhere else in momentum and energy.   As shown in classical, and collinear, Rb$_{2}$MnF$_{4}$\cite{Huberman2005} and quantum CFTD~\cite{Christensen2007, Piazza2014}, this spectral weight draws from the elastic channel in localized magnetic systems and this is consistent with the fact that neutron diffraction data reports a strongly reduced ordered moment for the Fe$^{3+}$ site while not for the Fe$^{2+}$.  As illustrated in Fig \ref{fig:two_magnon} $(b)$ and given the kinematic conditions, the gap and energy range of the single magnon modes provide favorable conditions to observe the decay of the higher energy Fe$^{3+}$ excitations.  

Similar momentum and energy broadened continuum have been reported in quantum ($S={1\over 2}$)~\cite{Stone2006,Masuda2006}, itinerant magnets~\cite{Stock2015}, and triangular systems~\cite{Coldea2003,Dalidovich2006}.   However, the observation of such a strong continuum and decay processes in a classical high spin magnet is unusual given predictions that such cross sections should scale as $\sim1/S$~\cite{Igarashi1992}. Moreover, while two-magnon scattering reported in other magnets were observed with a very weak intensity ($\sim$ 6\% of the integrated one-magnon intensity in \cite{Huberman2005}), it should be emphatized in this case that the broad continuum intensity in RbFe$^{2+}$Fe$^{3+}$F$_{6}$ was of the same order of magnitude as the sharp single magnon intensity.
RbFe$^{2+}$Fe$^{3+}$F$_{6}$ is thus a unique case where charge ordering allows the coupling of non-collinear spins oriented 90$^{\circ}$ to each other and demonstrates that this multi magnon phenomenon is not constrained to purely quantum systems and extends to classical magnets.  Such cross sections may be observable in other high spin magnets where similar ``orthogonal" or noncollinear spin arrangements exist and may include the oxyselenides and oxysulfides~\cite{Zhao2013,McCabe2014,Stock2016}.    The spin and charge degrees of freedom in RbFe$^{2+}$Fe$^{3+}$F$_6$ are well separated in terms of iron sites and also energy scales of branches.  The multiparticle excitations may provide a means of coupling charge and spin degrees of freedom in RbFe$^{2+}$Fe$^{3+}$F$_6$ and similar coupling processes have been suggested in BiFeO$_{3}$~\cite{Ramirez2008,Ramirez2009} and low dimensional cuprates~\cite{Hess2004}.

In summary, we report the magnetic fluctuations in charge ordered RbFe$^{2+}$Fe$^{3+}$F$_6$.  The separation of different Fe$^{2+}$ and Fe$^{3+}$ chains results in an orthogonal spin arrangement on the two different magnetic sites and separate spin-wave branches.  We observe multi magnon processes in this magnet and show that such processes can occur in classical magnets with a noncollinear spin arrangement.

\begin{acknowledgments}
We acknowledge funding from the EPSRC, STFC, and the Carnegie Trust for the Universities of Scotland.  We are thankful to E. Cussen (Strathclyde) for fruitful discussion.
\end{acknowledgments}

\bibliography{ms}

\end{document}


\title{Supplementary information for "Single and multi magnon excitations in charged-ordered RbFe$^{2+}$Fe$^{3+}$F$_6$"}

\author{M. Songvilay}
\affiliation{School of Physics and Astronomy and Centre for Science at Extreme Conditions, University of Edinburgh, Edinburgh EH9 3FD, UK}
\author{E. E. Rodriguez}
\affiliation{Department of Chemistry and Biochemistry, University of Maryland, College Park, Maryland 20742, USA}
\author{R. Lindsay}
\affiliation{School of Physical Sciences, University of Kent, Canterbury, CT2 7NH, UK}
\author{M. A. Green}
\affiliation{School of Physical Sciences, University of Kent, Canterbury, CT2 7NH, UK}
\author{H. C. Walker}
\affiliation{ISIS Neutron and Muon Source, Rutherford Appleton Laboratory, Chilton, Didcot OX11 0QX, United Kingdom}
\author{J. A. Rodriguez-Rivera}
\affiliation{NIST Center for Neutron Research, National Institute of Standards and Technology, 100 Bureau Drive, Gaithersburg, Maryland, 20899, USA}
\affiliation{Department of Materials Science, University of Maryland, College Park, Maryland 20742, USA}
\author{C. Stock}
\affiliation{School of Physics and Astronomy and Centre for Science at Extreme Conditions, University of Edinburgh, Edinburgh EH9 3FD, UK}
\date{\today}

\maketitle

\section{Materials synthesis}

RbFe$_{2}$F$_{6}$ was prepared hydrothermally using Parr Instrument Company general purpose acid digestion vessels with an internal volume of 45ml, 0.238 g (2.28 $\times$10$^{-3}$ mol) of RbF, 0.2141 g (2.28 $\times$10$^{-3}$ mol) of FeF$_{2}$ and 0.2573 g (2.28$\times$10$^{-3}$ mol) of FeF$_{3}$ were added to 3 ml (3.90 $\times$ 10$^{-2}$ mol) of CF$_{3}$COOH, and 5ml of H$_{2}$O in the autoclave. The autoclave was sealed and gradually heated to 230 $^{\circ}$C, held for 24 h before being cooled to room temperature at a rate of 0.1 $^{\circ}$C/min. The mother liquid was decanted and the remaining solid washed with H$_{2}$O using gravity filtration. The sample was then dried for approx. 12 h at 70 $^{\circ}$C before analysis.

\section{Neutron experiments}

Time-of-flight neutron spectroscopy measurements were performed using the MERLIN chopper spectrometer (ISIS, UK).  By using a gadolinium fermi chopper spinning at 200 Hz in multirep mode, incident energies of E$_{i}$=25 meV and 10 meV were simultaneously used.  Further triple-axis measurements were performed on the MACS cold triple-axis (NSIT, Gaithersburg) with E$_{f}$=3.7 meV and a BeO filter between the sample and the analyzer to remove higher order contamination.  Given the one dimensional nature, we utilized a configuration in both experiments where the incident beam is perpendicular to the chain $b$-axis allowing the multidetector arrangements to facilitate a scan along the $k$ direction.  

A rocking scan around the (0~2~0) reflection performed on MACS and presented in fig.\ref{fig:mosaic} shows that despite the co-alignment of many crystals, the mosaicity of the sample along the $k$ direction is 2.8$^{\circ}$, extracted from the a fit to a Gaussian function (full width at half maximum).

\setcounter{figure}{4}
\begin{figure}
 \includegraphics[scale=0.4]{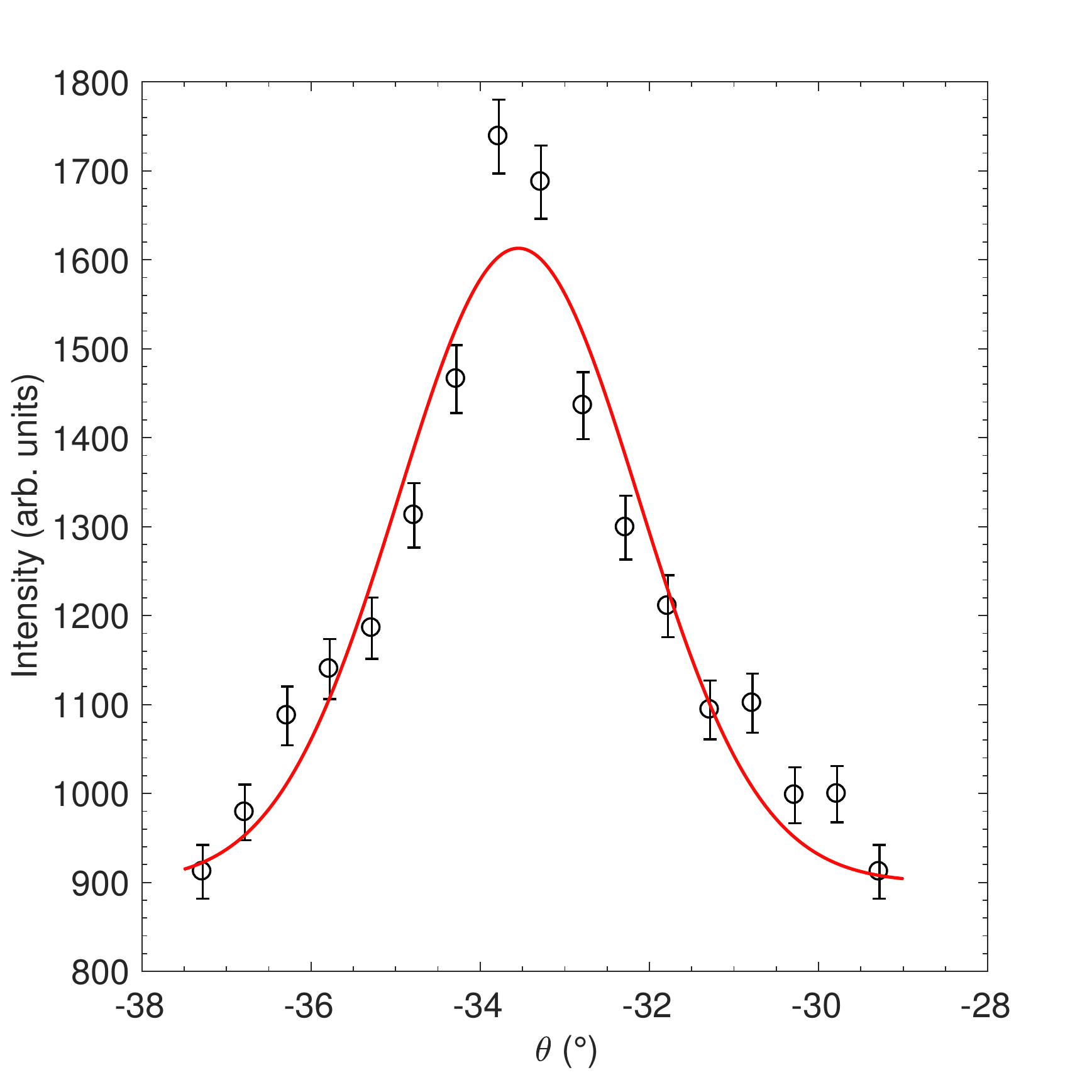}
 \caption{ \label{fig:mosaic} Rocking scan around the (0~2~0) reflection performed on MACS. The red line is a fit to a Gaussian function.}
\end{figure}

\section{Spinwave calculations - Spin Hamiltonian}
Based on the diffraction study, considering four different Fe-Fe distances and fours Fe-F-Fe angles, the nearest-neighbor Hamiltonian used to model the system is the following:

$$\mathcal{H}=\mathcal{H}_{nn}  + \mathcal{H}_{aniso}$$
Where 
\begin{eqnarray}
H_{nn} & = &\sum_{i,j,k} 
	J_{ch1} \mathbf{S_{ijk}^{1a}} \cdot (\mathbf{S_{ijk}^{1b}} + \mathbf{S_{(i+1)jk}^{1b}}) \nonumber\\*
&& + J_{ch1} \mathbf{S_{ijk}^{1c}} \cdot (\mathbf{S_{ijk}^{1d}} + \mathbf{S_{(i+1)jk}^{1d}}) \nonumber \\*
&& +  J_{ch2} \mathbf{S_{ijk}^{2a}} \cdot (\mathbf{S_{ijk}^{2b}} + \mathbf{S_{i(j-1)k}^{2b}}) \nonumber \\*
&& + J_{ch2} \mathbf{S_{ijk}^{2c}} \cdot (\mathbf{S_{ijk}^{2d}} + \mathbf{S_{i(j-1)k}^{2d}})  \nonumber \\*
&& + J_{int1}\mathbf{S_{i(j-1)k}^{1d}} \cdot (\mathbf{S_{i(j+1)k}^{2c}} + \mathbf{S_{i(j-1)k}^{2d}})\nonumber \\*
&& + J_{int1}\mathbf{S_{ijk}^{1a}} \cdot (\mathbf{S_{(i+1)(j+1)(k+1)}^{2c}} + \mathbf{S_{(i+1)(j+1)(k+1)}^{2d}})\nonumber \\*
&& + J_{int1}\mathbf{S_{ijk}^{2a}} \cdot (\mathbf{S_{i(j-1)k}^{1c}} + \mathbf{S_{ijk}^{1b}})\nonumber \\*
&& + J_{int1}\mathbf{S_{ijk}^{2b}} \cdot (\mathbf{S_{ijk}^{1c}} + \mathbf{S_{ijk}^{1b}})\nonumber \\*
&& + J_{int2}\mathbf{S_{(i+1)(j+1)k}^{1c}} \cdot (\mathbf{S_{(i+1)(j+1)k}^{2c}} + \mathbf{S_{(i+1)jk}^{2d}})\nonumber \\*
&& + J_{int2}\mathbf{S_{ij(k+1)}^{1b}} \cdot (\mathbf{S_{ij(k+1)}^{2c}} + \mathbf{S_{ij(k+1)}^{2d}})\nonumber \\*
&& + J_{int2}\mathbf{S_{ijk}^{2a}} \cdot (\mathbf{S_{ijk}^{1a}} + \mathbf{S_{i(j-1)k}^{1d}})\nonumber \\*
&& + J_{int2}\mathbf{S_{ijk}^{2b}} \cdot (\mathbf{S_{ijk}^{1d}} + \mathbf{S_{ijk}^{1a}})\nonumber \\*
\end{eqnarray}

and the Hamiltonian describing the anisotropy is:

\begin{equation}
H_{aniso}=\sum_{i,\alpha,\beta}D_y^{\alpha}( \mathbf{S_{i}^{\alpha}})^{2} + D_x^{\beta}( \mathbf{S_{i}^{\beta}})^{2}
\end{equation}

$\mathbf{S^{\alpha}_{ijk}}$ represents a spin on the Fe$^{2+}$ site where $\alpha \in \lbrace 1a, 1b, 1c, 1d \rbrace$, and $\mathbf{S^{\beta}_{ijk}}$ represents a spin on the Fe$^{3+}$ site where $\beta \in \lbrace 2a, 2b, 2c, 2d \rbrace$ as indicated in fig. 1 (a).  \textit{i}, \textit{j} and \textit{k} are the indices of the unit cell along the \textit{a}, \textit{b} and \textit{c} axes respectively. Two different octahedral environments for the two Fe sites are also taken into account, with an isotropic environment for the Fe$^{3+}$ site, and a significantly distorted environment for the Fe$^{2+}$ site. As a consequence, $D_y^{\alpha}$ and $D_x^{\beta}$ denote two different easy-axis anisotropy terms, for the Fe$^{3+}$ site and for the Fe$^{2+}$ site, respectively.

\subsection{Exchange couplings}
The best fit to the data using the spin wave model described above allowed to extract the exchange couplings summarized in table \ref{tab:couplings}. These values are also compared to Fe-Fe distances and Fe-F-Fe angles.

\begin{table*}[t]
\caption{\label{tab:couplings}Fe-Fe distances and angles at 4 K and the corresponding magnetic exchange couplings for RbFe$_2$F$_6$.}
\begin{ruledtabular}
\begin{tabular}{lllll}
& ch1& ch2 & int1 & int2\\
 \hline
 Fe-Fe distance ($\AA$)  & 3.5062(18) & 3.71952(2) & 3.5823(11) & 3.6460(11) \\
  J (meV) & 1.40(5) & 1.90(2) & 1.40(5) & 0.75(10)\\
 \hline
 & Fe$_1$-F-Fe$_1$ & Fe$_2$-F-Fe$_2$ & Fe$_1$-F-Fe$_2$ & Fe$_1$-F-Fe$_2$\\
 \hline
 Angles ($^{\circ}$) & 124.34(13) & 145.51(3) & 124.19(9) & 129.70(9) \\
\end{tabular}
\end{ruledtabular}
\end{table*}

\subsection{Spin wave excitations perpendicular to $b$}

Figure \ref{fig:mapL} displays the magnetic excitations measured perpendicularly to the $b$-axis on MERLIN with E$_i$ = 10 meV. The data, integrated along $k$ around 1 $\pm$ 0.25 (fig. \ref{fig:mapL} $(a)$) and 0.5 $\pm$ 0.25 (fig. \ref{fig:mapL} $(c)$) show very slightly dispersive modes along the $l$ direction. This indicates that the Fe$^{3+}$ and Fe$^{2+}$ chains are weakly coupled along the $c$ axis.  The spin wave calculations along $(0~1~l)$ and $(0~0.5~l)$ (fig. \ref{fig:mapL} $(b)$ and $(d)$ are displayed on the right panels and show an rather good agreement with the experimental data, although some discrepancies can be observed along $(0~0.5~l)$. The spin wave model was then used to calculate the excitation spectra in the $(h~1~0)$ and $(h~1.5~0)$ directions (fig. \ref{fig:mapL} $(f)$ and $(h)$). The calculations along $(h~1~0)$ show a flat mode around 2 meV and a dispersive higher energy mode around 4 meV, as well as 3 dispersive modes along $(h~1.5~0)$. The data in the $h$ direction were not recorded in a large Q range but shows an agreement with the calculations: figures \ref{fig:mapL} $(e)$-$(h)$ display a comparison between experimental data and calculations in the same Q range and the same number of modes with the right energy range can be distinguished in both directions. This hence confirms that the systems can be viewed as two weakly coupled Fe$^{3+}$ and Fe$^{2+}$ chains, despite the interchain couplings being less than an order of magnitude smaller than the intrachain couplings.
 
\setcounter{figure}{5}
\begin{figure}
 \includegraphics[scale=0.42]{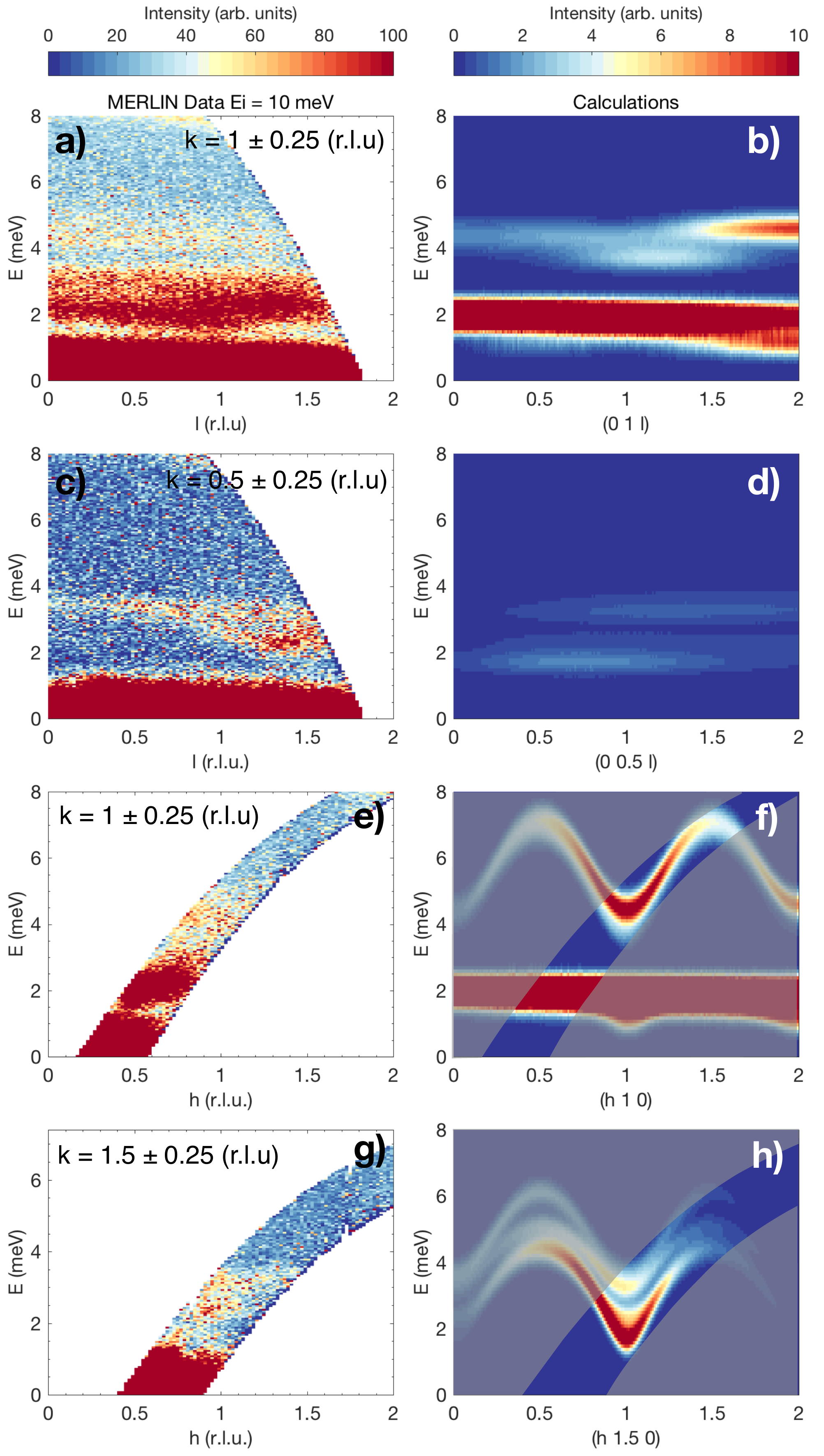}
 \caption{ \label{fig:mapL}  $left~panels$: Inelastic neutron scattering data of RbFe$_2$F$_6$ measured on MERLIN with incident energy of E$_i$ = 10 meV at T = 5 K, along $l$. $(a)-(d)$ and $h$ ($(e)-(g)$). $right~panels$: S($\mathbf{Q}$, $\omega$) simulated with spinwave calculations along $(0~1~l)$, $(0~0.5~l)$, $(h~1~0)$ and $(h~1.5~0)$. The grey areas on the $(f)$ and $(h)$ panels correspond to the regions in momentum and energy that were not covered by the experiment.}
\end{figure}